\documentclass[12pt,fleqn]{article}

\usepackage{amsmath,amsfonts,bm,amssymb}
\usepackage{amsthm}

\usepackage{ifpdf}

\usepackage{eucal}
\usepackage{eufrak}

\usepackage[margin=2cm,a4paper]{geometry}

\usepackage[round, sort&compress]{natbib}
\usepackage{url}

\usepackage{multirow}
\usepackage{rotating}

\usepackage{graphicx}

\usepackage{grffile}

\usepackage{color}
\usepackage[english]{babel}

\renewcommand{\leq}{\leqslant}

\newcommand{\remark}[1]{\marginpar{\scriptsize#1}}
\renewcommand{\remark}[1]{\marginpar{}}   

\long\def\symbolfootnote[#1]#2{\begingroup\def\thefootnote{\fnsymbol{footnote}}\footnote[#1]{#2}\endgroup}

\newcommand{\captionfonts}{\footnotesize}

\makeatletter  
\long\def\@makecaption#1#2{%
  \vskip\abovecaptionskip
  \sbox\@tempboxa{{\captionfonts #1: #2}}%
  \ifdim \wd\@tempboxa >\hsize
    {\captionfonts #1: #2\par}
  \else
    \hbox to\hsize{\hfil\box\@tempboxa\hfil}%
  \fi
  \vskip\belowcaptionskip}
\makeatother   

\newcommand{\be}{\begin{equation}} \newcommand{\ee}{\end{equation}}

\newcommand{\ben}{\begin{enumerate}} \newcommand{\een}{\end{enumerate}}
\newcommand{\bc}{\begin{center}} \newcommand{\ec}{\end{center}}
\newcommand{\bi}{\begin{itemize}} \newcommand{\ei}{\end{itemize}}

\newcommand{\cfm}{\mathrm{CFM}}
\newcommand{\mkt}{\mathrm{mkt}}

\begin{document}
 
\title{The short-term price impact of trades is universal}

\author{Bence T\'oth$^{\textrm{a}}$\footnote{Corresponding author. E-mail: ecneb.htot@gmail.com. Tel.: +33 1 49 49 59 49}~,
Zolt\'an Eisler$^{\textrm{a}}$\footnote{E-mail: zoltan.eisler@cfm.fr. Tel.: +33 1 49 49 59 49}~,
Jean-Philippe Bouchaud$^{\textrm{a}}$\footnote{E-mail: jean-philippe.bouchaud@cfm.fr. Tel.: +33 1 49 49 59 49}}
\maketitle
\small
\begin{center}
$^\textrm{a}$~\emph{Capital Fund Management, 23 Rue de l'Universit\'e, 75007 Paris, France}
\end{center}
\normalsize

\vspace{0.3cm}

\begin{abstract}
We analyze a proprietary dataset of trades by a single asset manager, comparing their price impact with that of the trades of the rest of the market. 
In the context of a linear propagator model we find no significant difference between the two, suggesting that both the magnitude and time dependence of impact are universal in anonymous, electronic markets. 
This result is important as optimal execution policies often rely on propagators calibrated on anonymous data. We also find evidence that in the wake of a trade 
the order flow of other market participants first adds further copy-cat trades enhancing price impact on very short time scales. The induced order flow then quickly inverts, thereby contributing to impact decay.
\end{abstract}

\smallskip


\clearpage


\section{Introduction}
\label{intro}

Trading impacts prices -- this is an undisputable empirical statement. However, the interpretation of this observed impact is still debated. Is it observed because trades {\it forecast} 
future price changes that would have happened anyway, as Efficient Market theorists would argue? Or should one better think of markets as a kind of physical medium that 
reacts statistically similarly to all trades, whether informed or not? 

This question is key to assess the relevance of most empirical studies of impact, that are based on anonymized order flow, where all orders are {\it de facto} treated on an equal footing. 
The average impact extracted from these studies could be misleading if ``informed'' and ``uninformed'' trades have completely different impacts. This is especially relevant  
in the context of popular models of impact, such as Hasbrouck's VAR model \citep{hasbrouck1991measuring, hasbrouck2007empirical} and the ``propagator'' model \citep{bouchaud2004subtle}, where the average reaction to market orders is deduced from a joint analysis of the order flow and 
the price time series (see also \citep{eislermodels,taranto2017,patzelt2017} for more recent developments). The standard practice is indeed to calibrate such models using public time series with anonymous orders, and deduce the average lag-dependent impact 
(or propagator) of market orders, with the hope of using this information to anticipate one's own market impact in future trading, and of timing executions accordingly \citep{gatheral2012transient}. 

In this study -- to our knowledge the first of its kind -- we aim to address this question by comparing the impact of the trades of Capital Fund Management (CFM) based on a proprietary dataset with the impact of the rest of the market. Perhaps surprisingly, we find no significant difference between the two. Our results strongly support the ``physical medium'' picture of markets, that statistically reacts to all perturbations in a similar way. Interestingly, we find that the reaction is first of a ``copy-cat'' type, where the initial trade is imitated by other market participants on very short time scales, while on longer time scales liquidity refill and/or trades in the opposite direction lead to impact reversion. On much longer time scales, the information motivating the trade (if any) is progressively revealed.

\section{Data and notations}
\label{data}

In the following we will analyze data on 98 liquid US stocks in the period 2 January 2014--31 December 2015. This includes all trades executed in the major lit markets. 
Each trade will be classified via a tag variable $\theta$. Trades by the strategies managed by CFM will be denoted as $\theta = \cfm$, whereas those by all other market participants 
-- which we will consider as a single representative actor -- as $\theta = \mkt$.
We will work in  \emph{trade time}, meaning that the time step counter $t$ is incremented whenever a trade occurs on the stock being studied. All trades on the same stock, reported in the same millisecond, at the same price, in the same direction and bearing the same participant tag are aggregated into one single transaction, because these usually correspond to the same single market order. After aggregation our dataset includes close to 520M trades, of which approx. $0.6 \%$ belong to CFM.

\begin{figure}[tb]
\begin{center}
  \includegraphics[scale=0.45]{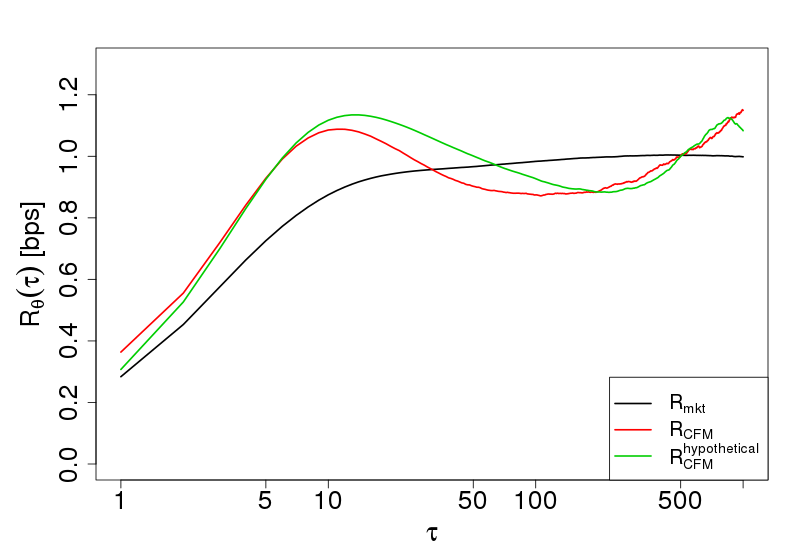}
\caption{The price response $\mathcal{R}_\theta$ for $\theta=\mkt$ (black) and $\theta=\cfm$ (red), averaged over all stocks in our pool. $\mathcal{R}_\cfm$ is stronger than $\mathcal{R}_\mkt$ for very short
times, but $\mathcal{R}_\cfm$ reverts  for $10<\tau<100$ and trends afterwards. The green line corresponds to the {\it reconstructed} response function, assuming that the single trade impact is the same for CFM and for the rest of the market, and accounting for a progressive revelation of the information content of the CFM trades. The final dip in the reconstructed response is a numerical artefact.} 
\label{fig:resp}
\end{center}
\end{figure}

We will denote by $\epsilon_t$ the side of the market order at time $t$, $\epsilon_t = +1$ for buyer initiated, and $-1$ for seller initiated ones. $m_t$ will be the mid price just before the trade $t$. We will use the indicator function $\mathbb{I}(A)$ defined as $\mathbb{I}(A) = 1$ if $A$ is true and $\mathbb{I}(A) = 0$ otherwise. For example, the expression $\mathbb{I}(\theta_t = \theta)$ is $1$ if the trade at time $t$ resulted from a market order by $\theta$. The unconditional probability of the type $\theta$ is thus by definition $P(\theta) = \langle \mathbb{I}(\theta_t = \theta)\rangle$. Finally, as discussed in Ref. \citep{eislermodels}, for any quantity $X$ its average sampled at trades initiated by an actor of type $\theta$ is
\be
  \langle X \vert \theta_t=\theta \rangle = \frac{\langle X \mathbb{I}(\theta_t=\theta)\rangle}{P(\theta)}.
\ee

\section{Price impact and order flow with labeled data}

Let us define the \emph{response function} of trades \citep{bouchaud2004subtle, toth2012does} initiated by actor $\theta$ as the average subsequent change of the mid price in the direction of the trade:
\be
\mathcal{R}_\theta(\tau)=\frac{\langle (m_{t+\tau}-m_t) \mathbb{I}(\theta_t=\theta)\epsilon_t\rangle}{P(\theta)}.
\ee
The average of this quantity across all stocks in our dataset is shown in Fig. \ref{fig:resp}. The overall shape of $\mathcal{R}_\mkt(\tau)$ is compatible with that previously reported in the literature. On the other hand, $\mathcal{R}_\cfm(\tau)$ is clearly different, with a hump for short times, and then a continuous upward trend. We will discuss the possible origin of these differences in the following.

\begin{figure}[tb]
\begin{center}
  \includegraphics[scale=0.45]{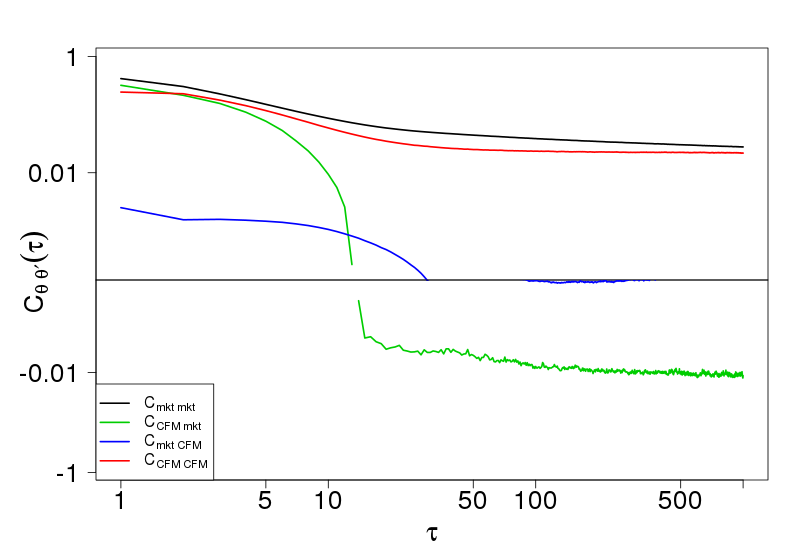}
\caption{The correlation functions, $C_{\theta,\theta^\prime}$, for 
$\theta^\prime=\mkt$ and $\theta=\mkt$ (black),
$\theta^\prime=\mkt$ and $\theta=\cfm$ (green),
$\theta^\prime=\cfm$ and $\theta=\mkt$ (blue),
$\theta^\prime=\cfm$ and $\theta=\cfm$ (red),
averaged over all stocks in our pool.
Note, that the vertical axis is logarithmic both in the positive and negative planes.
We see that while $C_{\mkt,\mkt}$ decays
slowly and remains positive up to very long times, $C_{\cfm,\mkt}$ becomes rapidly negative after the initial positive part, and remains weakly negative
afterwards.}
\label{fig:corr}
\end{center}
\end{figure}

It is convenient to define the correlation function of the market order flow corresponding to two tags $\theta$ and $\theta^\prime$ as
\be
C_{\theta,\theta^\prime}(\tau > 0)=\frac{\langle \mathbb{I}(\theta_t=\theta)\epsilon_t \mathbb{I}(\theta_{t+\tau}=\theta^\prime)\epsilon_{t+\tau}\rangle}{P(\theta)P(\theta^\prime)},
\ee
where conventionally $\theta$ is the first to trade. Considering that both $\theta$ and $\theta^\prime$ can be $\cfm$ or $\mkt$, there are four possible combinations, 
that are shown in Fig. \ref{fig:corr}. The market-market order flow autocorrelation $C_{\mkt,\mkt}(\tau)$ is well known 
to be positive and slowly decaying, as a consequence of order splitting \citep{lillo2004long,toth2015why}. Since liquidity at the best bid and ask in order books is typically small, market participants slice their full quantity they intend to trade into correspondingly small orders. This results in the observed persistency of order flow revealed by $C_{\mkt, \mkt}$.
$C_{\cfm, \cfm}$ is qualitatively similar to $C_{\mkt, \mkt}$, decaying slowly.

The correlation $C_{\cfm, \mkt}(\tau)$, on the other hand, measures how the market ``reacts'' to CFM's trades. For short time lags we observe a strong positive correlation, i.e. other actors herd on CFM's trade. This correlation then quickly reverts on the scale of $\sim 10$ trades and becomes negative for longer time lags, indicating a contrarian behaviour of the rest of the market -- in other words, a flow of buy orders generates a counter-flow of sell orders, as discussed in many papers (see e.g. \citep{bouchaud2006random, weber, eislermodels, toth2012does}).
We also show the correlation $C_{\mkt, \cfm}(\tau)$, measuring how CFM reacts to the order flow of the market. This correlation is two orders of magnitude weaker than the 
others.

\begin{figure}[tb]
\begin{center}
  \includegraphics[scale=0.45]{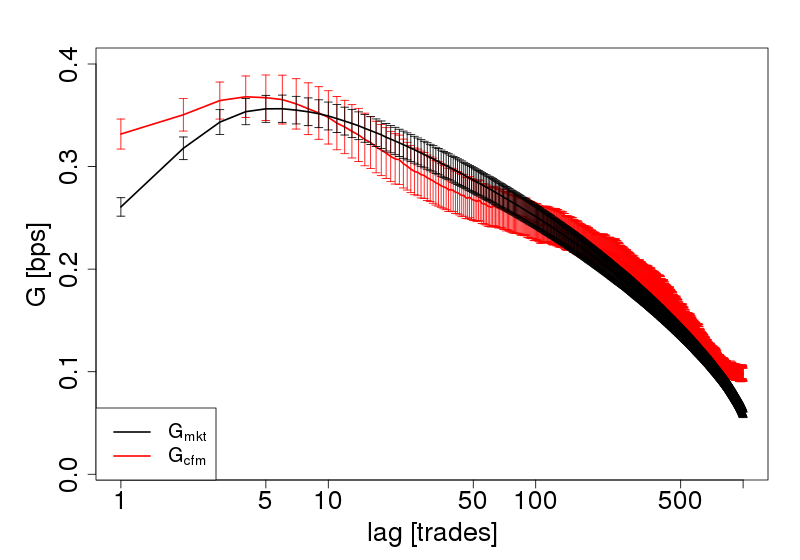}
\caption{The propagators, $G_\theta$, for $\theta=\mkt$ (black) and $\theta=\cfm$ (red). Error bars are estimated by bootstrapping over the pool of stocks. Apart from the difference for very short times, the two curves are very similar. The data for $\tau \gtrsim 500$ is affected both by the (numerical) boundary conditions at $\tau=1000$, and by the progressive revelation of the information content of CFM's trades.}
\label{fig:G}
\end{center}
\end{figure}

In order to more precisely quantify the causal relationship between individual orders and subsequent price changes, it is customary \citep{bouchaud2004subtle, eislermodels} to assume that 
the price response to market orders can be described by linear kernels $G_\theta$, i.e.:
\be
m_t=m_{-\infty}+\sum_{\theta}\sum_{t'<t}G_\theta(t-t')\mathbb{I}(\theta_{t'}=\theta)\epsilon_{t'}+\sum_{t'<t}\zeta_t,
\ee
where $\zeta_t$ is a zero mean, i.i.d. noise term. In this equation the {\it propagator} $G_\theta(t-t')$ represents the evolution of the mid price from $t'$ to $t$ given that the actor $\theta$ executed an isolated buy order in $t'$. The model can be calibrated using the following relation between $\mathcal{R}$, $C$ and $G$ \citep{eislermodels}:
\begin{eqnarray}
    \mathcal{R}_{\theta}(\tau+1)-\mathcal{R}_{\theta}(\tau) = \sum_{\theta^\prime} P(\theta^\prime) \left[ G_{\theta^\prime}(0) C_{\theta, \theta^\prime}(\tau) + \sum_{n>-\tau} \left[G_{\theta^\prime}(\tau+n+1)
    - G_{\theta^\prime}(\tau+n)\right] C_{\theta, \theta^\prime}(n) \right].
    \label{eq:undresspi}
\end{eqnarray}
We have injected the average $\mathcal{R}_\theta$'s and $C_{\theta, \theta^\prime}$'s plotted in Figs. \ref{fig:resp} and \ref{fig:corr}, for $1 \leq \tau \leq 1000$, to obtain the propagators $G_\mkt(\tau)$ and $G_\cfm(\tau)$. We show the results in Fig. \ref{fig:G}, together with error bars that are estimated by bootstrapping over the pool of stocks. We claim that apart from very short time scales, no significant difference between the effect of CFM's trades and those of the rest of the market. For short time scales, we attribute the remaining discrepancies to the fact that CFM volumes being on average larger than the rest of the market, while in our model all trades, independently of their volume, are assumed to impact the market in the same way. At large lags, there are systematic biases in the estimates of the $G$'s coming from boundary effects (i.e., implicitly setting $G_{\theta}(\tau)=$ cst. for $\tau > 1000$), so that the results cannot be trusted in that region. Furthermore, the information content of trades is expected to progressively show up. 

In order to bolster our claim that $G_\mkt(\tau) \approx G_\cfm(\tau)$, we have computed the reconstructed response function to CFM's trades, using Eq. (\ref{eq:undresspi}) with $G_\cfm(\tau)=G_\mkt(\tau)$, and adding an information revelation contribution proportional to $\tau$ itself, leaving the slope as an extra fitting parameter. As can be seen in Fig. \ref{fig:resp}, the reconstructed response function reproduces very well all the different features of $\mathcal{R}_{\cfm}(\tau)$. Therefore, the difference between the response functions reported in Fig. \ref{fig:resp} can chiefly be attributed to different cross-correlations between order flows and information revelation, but does not come from the ``bare'' reaction of the market encoded in the propagators. This is the central message of our paper.  

Let us stress that the curves in Fig. \ref{fig:G} represent an average over all 98 stocks in our sample, so one could wonder whether the agreement reported here still holds for each individual stock. The results are obviously much more noisy for the individual stocks separately, but the conclusion is the same, i.e. there is no systematic difference between the time dependence of the market propagator $G_\mkt(\tau)$ and the CFM propagator $G_\cfm(\tau)$. The global scale of these propagators can, however, be different for certain stocks. This is related to the fact that the average size of CFM's individual trades can be different from that of the rest of the market.   

\section{Cross-correlations and induced trades}

In order to dwell further on the effects induced by order flow correlations, we deconvolute the correlation functions $C_{\theta,\theta^\prime}$ in terms of flow-reaction kernels 
$K_{\theta,\theta^\prime}$ defined through the following regression \citep{eislermodels}:
\be
\mathbb{I}(\theta_t=\theta)\epsilon_t=\sum_{\theta^\prime}\sum_{t'<t}K_{\theta^\prime,\theta}(t-t')\mathbb{I}(\theta_{t'}=\theta^\prime)\epsilon_{t'}+\mu_t,
\ee
where $\mu_t$ is again a zero mean, i.i.d. noise. Here $K_{\theta^\prime, \theta}(t-t')$ is the expected order flow imbalance (probability of buys minus probability of sells) of actor $\theta$ at time $t$ conditional to actor $\theta^\prime$ sending a buy order at an earlier time $t'$.

\begin{figure}[tb]
\begin{center}
  \includegraphics[scale=0.45]{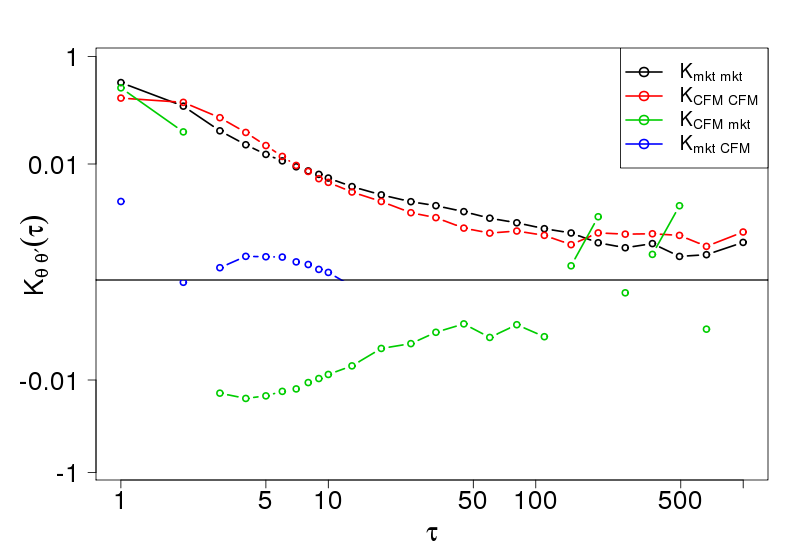}
\caption{The flow-reaction kernels $K_{\theta,\theta^\prime}$.
Note that the vertical axis is logarithmic both in the positive and negative planes.
We find that $K_{\cfm,\cfm}$ (red) and $K_{\mkt,\mkt}$ (black) are very similar. $K_{\cfm,\mkt}$ (green) is positive for very short times and becomes persistently negative later.
$K_{\mkt,\cfm}$ (blue) is much weaker than the other terms.}
\label{fig:K}
\end{center}
\end{figure}

Similarly to correlation functions there are four such kernels, shown in Fig. \ref{fig:K}. The self-kernels $K_{\mkt, \mkt}$ and $K_{\cfm, \cfm}$ are quite similar. From the literature and our own experience we know that both curves are dominated by order splitting, common to most large investors \citep{lillo2004long,toth2015why}. Note however, that $K_{\mkt, \mkt}$ contains both a contribution from order splitting and a contribution from reaction trades among other participants. The kernel $K_{\cfm, \mkt}$ -- describing purely how the market reacts to CFM's trades -- is quite interesting. Its shape supports what was expected based on the sign correlation itself. Notably, on a short time scale (in this case up to lag 2) trades by CFM induce copy-cat trades by the market. This effect then reverts and gives way to a weaker, but much more persistent (up to at least 100 trades) flow of contrarian orders by others. This direct observation of the interaction of aggressive order flows from different actors is another noteworthy result of our study. It is also clearly related to the previously documented passive limit order counter-flow (``liquidity refill''), which manifests itself as the decay of the propagator $G$ \citep{bouchaud2006random, eislermodels, toth2012does}.

\section{The total impact of a single trade}

\begin{figure}[tb]
\begin{center}
  \includegraphics[scale=0.45]{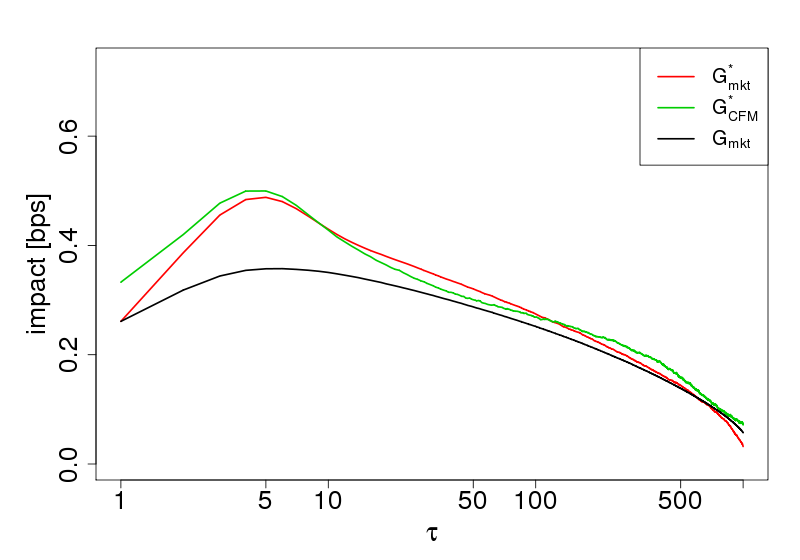}
\caption{Comparison of $G_\mkt$ (black), $G^\star_{\mkt}$ (red) and $G^\star_{\cfm}$ (green).
The total impact of a single trade is enhanced at short times by copy-cat trades, but afterwards decays back close to the bare propagator $G$.
$G^\star_{\cfm}$ is very similar to $G^\star_{\mkt}$ in line with our 
assumption that CFM trades are not statistically different from those of the market.
}
\label{fig:Gstar}
\end{center}
\end{figure}

By definition, the impact of an isolated trade, stripped off from its ``cloud'' of correlated trades, is measured by the propagators $G_\mkt(\tau)$, $G_\cfm(\tau)$. 
However, we now know that even if a given market participant chooses to execute a single trade, it will be followed by other ``reaction'' trades, with a probability encoded by the 
cross kernel $K$ \citep{eislermodels}. The total unconditional impact, denoted $G^\star_\cfm(\tau)$, of a single market order sent by CFM is therefore given by the sum of the single trade impact and the impact of the induced trades, namely,
\be\label{eq:Gstar}
G^\star_{\cfm}(\tau)=G_{\cfm}(\tau)+\sum_{0<\tau_1<\tau} K_{\cfm,\mkt}(\tau_1) G^\star_{\mkt}(\tau-\tau_1) ,
\ee
where $G^\star_\mkt(\tau)$ is the total market impact, itself dressed by its own reaction trades. The latter can be represented as
\be
G^\star_{\mkt}(\tau)=G_{\mkt}(\tau)+\sum_{0<\tau_1<\tau}K_{\mkt,\cfm}(\tau_1) G^\star_{\cfm}(\tau-\tau_1)+\sum_{0<\tau_1<\tau}K_{\mkt,\mkt'}(\tau_1) G^\star_{\mkt}(\tau-\tau_1),
\label{eq:herdingratio}
\ee
where we introduced a new function: $K_{\mkt,\mkt'}(\tau)$ describes the reaction to an order by a given market participant, by all other market participants {\it excluding CFM}. 
Exactly as the propagator for CFM trades is nearly identical to that of any market trade, it is reasonable to assume that in an anonymous market the same is true for the reaction of the order flow itself. 
In other words, one expects that
\be
K_{\mkt,\mkt'}(\tau) \approx K_{\cfm,\mkt}(\tau),
\label{eq:mktcfm}
\ee
possibly with an overall multiplicative factor that accounts for the difference in traded volumes. Furthermore, from the blue curve in Fig. \ref{fig:K} we see that the reaction of CFM's order flow to market trades is negligible, i.e. $K_{\mkt,\cfm}(\tau) \approx 0$. This final observation along with Eq. \eqref{eq:mktcfm} allows us to simplify Eq. \eqref{eq:herdingratio} to a closed, self-consistent equation for $G^\star_\mkt(\tau)$ in terms of observable quantities:
\be
G^\star_{\mkt}(\tau) \approx G_{\mkt}(\tau)+\sum_{0<\tau_1<\tau}K_{\cfm,\mkt}(\tau_1) G^\star_{\mkt}(\tau-\tau_1).
\label{eq:herdingratio2}
\ee
This equation can be solved numerically, leading to the result plotted in Fig. \ref{fig:Gstar}. As expected from the short-term herding effect unveiled in the previous section, we see that 
the total impact of a single trade is enhanced at short times by copy-cat trades, but afterwards decays back close to the bare propagator $G$.
Comparing Eq. \eqref{eq:herdingratio2} with Eq. \eqref{eq:Gstar} furthermore yields $G^\star_\cfm(\tau) \approx G^\star_\mkt(\tau)$, as a direct consequence of our assumption that 
CFM trades are not statistically different from those of the market.

\section{Conclusion}

Based on our proprietary dataset, we have argued that there is no significant difference, at the single trade level, between the short term price impact of CFM and that of the rest of the market. This is a strong indication of the universality of impact across all market participants, bolstering the idea that short term impact is a property of the market itself, and has no relation whatsoever to the information content of trades that only reveals itself on longer time scales (hundreds of trades in the present case). In an anonymous market we indeed expect the same collective behavior to follow the actions of any market participant. Our findings empirically support a ``physical medium'' theory of market impact, which is relevant in practice as optimal execution policies often implicitly rely on such an assumption. The interaction of aggressive order flow among actors contributes to price moves in a small, but significant way, and its calibration on anonymous data is warranted by the present results.


\section*{Acknowledgements} 

We thank X. Brokmann, J. Donier, M. Gould, J. Kockelkoren, I. Mastromatteo and in particular J. Bonart for many useful discussions and suggestions on the content of this paper. 

\bibliographystyle{abbrvnat}
\bibliography{cfm_vs_mkt}

\begin{thebibliography}{12}
\providecommand{\natexlab}[1]{#1}
\providecommand{\url}[1]{\texttt{#1}}
\expandafter\ifx\csname urlstyle\endcsname\relax
  \providecommand{\doi}[1]{doi: #1}\else
  \providecommand{\doi}{doi: \begingroup \urlstyle{rm}\Url}\fi

\bibitem[Bouchaud et~al.(2004)Bouchaud, Gefen, Potters, and
  Wyart]{bouchaud2004subtle}
J.-P. Bouchaud, Y.~Gefen, M.~Potters, and M.~Wyart.
\newblock Fluctuations and response in financial markets: the subtle nature of
  random price changes.
\newblock \emph{Quantitative {F}inance}, 4\penalty0 (2):\penalty0 176--190,
  2004.

\bibitem[Bouchaud et~al.(2006)Bouchaud, Kockelkoren, and
  Potters]{bouchaud2006random}
J.-P. Bouchaud, J.~Kockelkoren, and M.~Potters.
\newblock Random walks, liquidity molasses and critical response in financial
  markets.
\newblock \emph{Quantitative Finance}, 6\penalty0 (2):\penalty0 115--123, 2006.

\bibitem[Eisler et~al.(2013)Eisler, Bouchaud, and Kockelkoren]{eislermodels}
Z.~Eisler, J.-P. Bouchaud, and J.~Kockelkoren.
\newblock Models for the impact of all order book events.
\newblock \emph{Market Microstructure: Confronting Many Viewpoints}, pages
  113--135, 2013.

\bibitem[Gatheral et~al.(2012)Gatheral, Schied, and
  Slynko]{gatheral2012transient}
J.~Gatheral, A.~Schied, and A.~Slynko.
\newblock Transient linear price impact and fredholm integral equations.
\newblock \emph{Mathematical Finance}, 22\penalty0 (3):\penalty0 445--474,
  2012.

\bibitem[Hasbrouck(1991)]{hasbrouck1991measuring}
J.~Hasbrouck.
\newblock Measuring the information content of stock trades.
\newblock \emph{The Journal of Finance}, 46\penalty0 (1):\penalty0 179--207,
  1991.

\bibitem[Hasbrouck(2007)]{hasbrouck2007empirical}
J.~Hasbrouck.
\newblock \emph{Empirical Market Microstructure}.
\newblock Oxford University Press New York, 2007.

\bibitem[Lillo et~al.(2004)Lillo, Farmer, et~al.]{lillo2004long}
F.~Lillo, J.~D. Farmer, et~al.
\newblock The long memory of the efficient market.
\newblock \emph{Studies in {N}onlinear {D}ynamics \& {E}conometrics},
  8\penalty0 (3):\penalty0 1, 2004.

\bibitem[Patzelt and Bouchaud(2017)]{patzelt2017}
F.~Patzelt and J.-P. Bouchaud.
\newblock Nonlinear price impact from linear models.
\newblock \emph{Journal of Statistical Mechanics: Theory and Experiment},
  2017\penalty0 (12):\penalty0 123404, 2017.

\bibitem[Taranto et~al.(2016)Taranto, Bormetti, Bouchaud, Lillo, and
  T\'oth]{taranto2017}
E.~D. Taranto, G.~Bormetti, J.-P. Bouchaud, F.~Lillo, and B.~T\'oth.
\newblock Linear models for the impact of order flow on prices i. propagators:
  Transient vs. history dependent impact.
\newblock \emph{to appear in Quantitative Finance}, 2016.

\bibitem[T\'oth et~al.(2012)T\'oth, Eisler, Lillo, Kockelkoren, Bouchaud, and
  Farmer]{toth2012does}
B.~T\'oth, Z.~Eisler, F.~Lillo, J.~Kockelkoren, J.-P. Bouchaud, and J.~D.
  Farmer.
\newblock How does the market react to your order flow?
\newblock \emph{Quantitative Finance}, 12\penalty0 (7):\penalty0 1015--1024,
  2012.

\bibitem[T\'oth et~al.(2015)T\'oth, Palit, Lillo, and Farmer]{toth2015why}
B.~T\'oth, I.~Palit, F.~Lillo, and J.~D. Farmer.
\newblock Why is equity order flow so persistent?
\newblock \emph{Journal of Economic Dynamics and Control}, 51:\penalty0
  218--239, 2015.

\bibitem[Weber and Rosenow(2005)]{weber}
P.~Weber and B.~Rosenow.
\newblock Order book approach to price impact.
\newblock \emph{Quantitative Finance}, 5\penalty0 (4):\penalty0 357--364, 2005.

\end{thebibliography}

\end{document}